\documentclass[journal=nalefd,manuscript=letter,layout=onecolumn]{achemso}
\usepackage[colorlinks,linkcolor={blue},citecolor={blue},urlcolor={blue}]{hyperref} 
\usepackage{graphicx}
\usepackage{dcolumn}
\usepackage{bm}
\usepackage[mathlines]{lineno}
\usepackage{mathrsfs}
\usepackage{amsfonts}
\usepackage[T1]{fontenc}
\usepackage{float}
\newcommand{\bl}{\boldsymbol}
\usepackage{siunitx}
\usepackage{xcolor}
\usepackage{amssymb}
\usepackage{float}

\title{Topological flat bands in strained graphene: substrate engineering and optical control}

\author{Md Tareq Mahmud}
\affiliation{Physics and Astronomy Department, and Nanoscale and Quantum Phenomena Institute, Ohio University, Athens, Ohio 45701-2979, USA}
\alsoaffiliation{Present address: Department of Physics, University of Dhaka, Dhaka-1000, Bangladesh}
\altaffiliation{M.T.M. and D.Z. contributed equally to this work.}

\author{Dawei Zhai}
\email{dzhai@hku.hk}
\affiliation{Department of Physics, The University of Hong Kong, Hong Kong, China}
\alsoaffiliation{HKU-UCAS Joint Institute of Theoretical and Computational Physics at Hong Kong, Hong Kong, China}
\altaffiliation{M.T.M. and D.Z. contributed equally to this work.}

\author{Nancy Sandler}
\email{sandler@ohio.edu}
\affiliation{Physics and Astronomy Department, and Nanoscale and Quantum Phenomena Institute, Ohio University, Athens, Ohio 45701-2979, USA}

\begin{document}

\begin{abstract}
The discovery of correlated phases in twisted moir\'e superlattices accelerated the search for low-dimensional materials with exotic properties. A promising approach uses engineered substrates to strain the material. However, designing substrates for tailored properties is hindered by the incomplete understanding of the relationship between substrate's shapes and electronic properties of the deposited materials. By analyzing effective models of graphene under periodic deformations with generic crystalline profiles, we identify strong $C_{2z}$ symmetry breaking as the critical substrate geometric feature for emerging energy gaps and quasi-flat bands. We find continuous strain profiles producing connected pseudo-magnetic field landscapes are important for band topology. We show that the resultant electronic and topological properties from a substrate can be controlled with circularly polarized light, which also offers unique signatures for identifying the band topology imprinted by strain. Our results can guide experiments on strain engineering for exploring interesting transport and topological phenomena.
\end{abstract}


%%%%%%%%%%%%%%%%%%%%%%%%%%%%%%%%%%%%%%%%%%%%%%%%%%%%%%%%%%%%%%%%%%
Striking experiments have shown that bilayer graphene can become superconducting when appropriate twists between the layers are applied~\cite{Cao2018Supercond}. The phenomena is rooted in modifications of the interlayer hybridization, which is well described by effective Dirac models rendering flat bands for specific twist angles~\cite{McDonald2010,Tarnopolsky2019}. 

Strain has been suggested as an alternative to induce flat bands that favor the emergence of superconductivity or other correlated phases~\cite{Volovik2018}. 
A recent experiment has reported signatures of isolated flat bands and correlated states in buckled graphene on NbSe$_2$~\cite{Mao2020}. 
In addition to spontaneous buckling effects, periodic strain/deformations can also be achieved by 
thermal treatments~\cite{Bao2009Ther,Meng2013Ther,Bai2014Ther}, pressure~\cite{Bunch2008Pres,Zabel2012Pres,Boddeti2013Pres,Shin2016Pres}, mechanical manipulation~\cite{Klimov2012,Zhu2014,NemesIncze2017,Li2020Polymer}, and 
depositing the material on patterned substrates with controlled designs~\cite{Tomori2011Subs,Lu2012,Plantey2014Subs,Palinkas2016,Pacakova2017Subs,Jiang2017,Zhang2018Subs,Li2021a,Dai2019,Kollipara2020,YangS2021,Weinhold2021}.
In these scenarios, the emergent superlattice structure imposes new constraints on the electron dynamics akin to those produced in the twisted bilayer configuration.

Among the various approaches, substrate design holds the potential for the fabrication of large-scale samples with device applications. 
In addition, the samples can be easily investigated with local and non-local techniques under fixed strain configurations.
Experiments using substrates to induce periodic deformations on graphene membranes have reported unexpected electronic transport~\cite{Zhang2019,Jessen2019,Wu2022} and peculiar local density of states (LDOS) features~\cite{Banerjee2020,Hsu2020a}. 
These developments were complemented by theoretical studies of effective models, which treat the effects of strain in terms of a pseudo-magnetic field~\cite{Neek-Amal2012,Neek-Amal2014b,Milovanovic2019,Mahmud2020,Milovanovic2020a,Giambastiani2021,Phong2021a,DeBeule2022,Gao2022,Lin2023,Manesco2021,Manesco2021a}.
Although intriguing phenomena have been predicted, many of these works assume the pseudo-magnetic field profile proposed in the particular setup of buckled graphene on NbSe$_2$~\cite{Mao2020}.
One naturally expects that details of the pseudo-magnetic field should depend on the specific experimental setting, i.e., strain profiles. However, there is a gap in connecting electronic behavior to distinct strain profiles. Knowledge of the key geometric features of the patterned substrates that determine the resulting electronic properties is highly desirable for guiding future experiments.
Also, experimental evidence of the nontrivial band topology, which is predicted in periodically strained graphene and shown to be important for some correlation-driven phenomena~\cite{Phong2021a,DeBeule2022,Gao2022,Lin2023}, remains elusive.
Consequently, the potential for substrate design to produce tailored electronic and topological properties remains untapped.

We address these issues by analyzing graphene membranes subject to generic deformation profiles induced by substrates with different periodic shapes. 
By comparing substrate geometries and electronic structures, we determine strong $C_{2z}$ symmetry breaking as the geometric feature for large gap openings and the emergence of quasi-flat bands. From the charge distributions, we elucidate the crucial pseudo-magnetic field profile features that produce topological bands. 
These findings are experimentally accessible with scanning probes, transport measurements, and optical techniques.
Finally, we show how to control the electronic and topological properties using circularly polarized light, which also provides unique signatures for the experimental identification of the strain-induced band topology.

%%%%%%%%%%%%%%%%%%%%%%%%%%%%%%%%%%%%%%%%%%%%%%%%%%%%%%%%%%%

The substrates considered have periodicity on length scales larger than atomic-bond distance and produce smooth modulations on the membrane's height. 
Following common experimental practices,~\cite{Mao2020,Tomori2011Subs,Plantey2014Subs,Zhang2018Subs,Zhang2019,Jiang2017} we assume that the substrates are made of materials whose hybridization with graphene is negligible, such that the electronic properties of graphene are maintained. The major role of the substrate is to imprint periodic deformation/strain onto graphene, the effects of which will be studied in the following.
Under these conditions, electron dynamics can be captured by an effective Dirac model around graphene valleys $K$ and $K'$. The deformation is characterized by the strain tensor with Cartesian components $\epsilon_{ij}=\frac{1}{2} (\partial_{i}u_{j} + \partial_{j}u_{i}+\partial_{i}h\partial_{j}h)$, where $\bl{u}$ and $h$ denote in-plane and out-of-plane atomic displacements, respectively. In the continuum model, effects of $\epsilon_{ij}$ are represented by a pseudo-gauge potential $\bl{A_p}$~\cite{Vozmediano2010,Katsnelson2012}, and the Hamiltonian takes the form:
\begin{equation}
H_{\tau} = v\bl{\sigma}_{\tau}\cdot(\bl{p}+\tau e\bl{A_p}),
\label{eq:Ham}
\end{equation}
where $\tau=\pm$ refers to the $K/K'$ valley, $\bl{\sigma}_{\tau}=(\tau\sigma_{x},\sigma_{y})$ is composed of Pauli matrices acting on the sublattice spinor space, and $v$ is the Fermi velocity.
A small scalar potential may also exist, whose role is discussed in the Supporting Information (SI).
In terms of the strain tensor components, $\bl{A_p}=\frac{\hbar\beta}{2a}(\epsilon_{xx}-\epsilon_{yy}, -2\epsilon_{xy})$ with $\beta \approx 3$ and $a=1.42$~\AA. A proper spatial dependence gives rise to a pseudo-magnetic field $\tau\bl{B_p}=\tau\bl{\nabla}\times\bl{A_p}$ with opposite signs between valleys. The sign change ensures that $\tau\bl{B_p}$ preserves time-reversal symmetry (TRS) overall, although it breaks an effective TRS in each valley~\cite{Gusynin2007,Herbut2008}.

%%%%%%%%%%%%%%%%%%%%%%%%%%%%%%%%%%%%%%%%%%%%%%%%%

\begin{figure}[t]
	\includegraphics[width=6in]{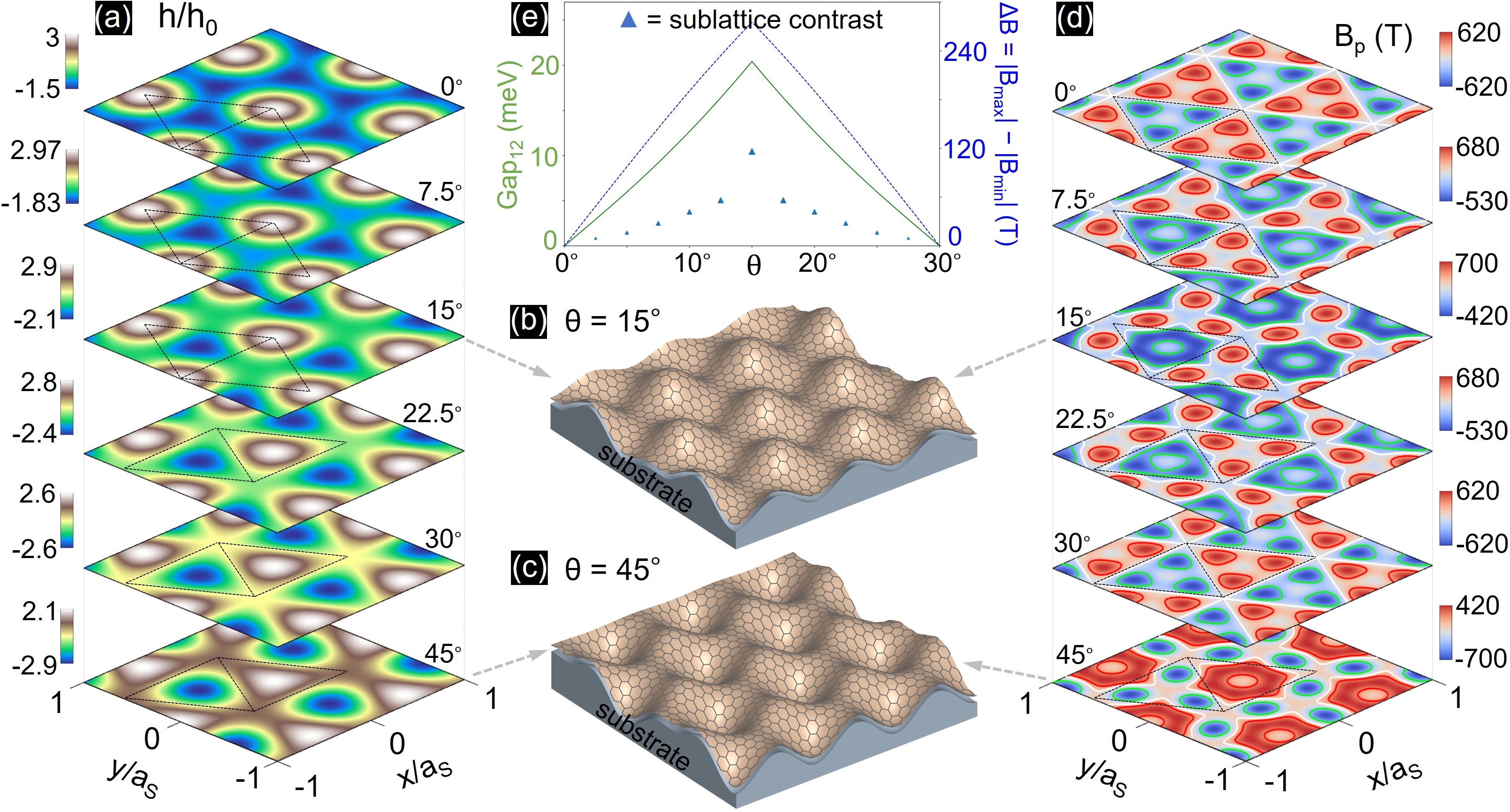}
	\caption{2D (a) and 3D (b, c) illustrations of height profiles, and 2D maps for $\bl{B_{p}}$ (d), induced by the deformation described by Eq.~(\ref{eq:height}). Dashed triangles in (a) and (d) define two `superlattice sublattice' sites, highlighting the $C_{2z}$ symmetry breaking for values of $\theta\ne0^\circ$ and $30^\circ$. (e) Variation of the first band gap (see Fig.~\ref{fig:Fig2}), sublattice LDOS contrast $(\rho_A^{\text{max}}-\rho_B^{\text{max}})/(\rho_A^{\text{max}}+\rho_B^{\text{max}})$, and $\bl{B_{p}}$ magnitude asymmetry as a function of $\theta$. Here $h_0 = 0.6$~nm, $a_S = 14$~nm (maximum strain $\simeq 18\%$).}
	\label{fig:Fig1}
\end{figure}

We illustrate our main findings with a model of graphene deposited on a substrate with protrusions arranged in a triangular superlattice as shown schematically in Figs.~\ref{fig:Fig1}(b) and (c). Other lattice geometries exhibit similar properties (see SI). In all the envisioned geometries, out-of-plane deformations introduce the main effects, while contributions from in-plane displacements add high-order corrections (see SI). Such details do not affect the qualitative features discussed below. The superlattice is oriented with its $\bl{x}$-axis along the zigzag direction of graphene, and the superlattice constant $a_{S} \gg a$ is chosen to be commensurate with graphene's~\cite{Phong2021a}. With these conventions, the height of the substrate surface is modeled by: 
\begin{equation}
h=h_{0}\displaystyle\sum_{i=1,\,2,\,3}\cos(\bl{b}_{i}\cdot\bl{r}+\theta),
\label{eq:height}
\end{equation}
where $\bl{b}_{1}=(1,-\frac{1}{\sqrt{3}})\frac{2\pi}{a_S}$, $\bl{b}_{2}=(0,\frac{2}{\sqrt{3}})\frac{2\pi}{a_S}$, and $\bl{b}_{3}=-\bl{b}_{1}-\bl{b}_{2}$ define a superlattice with a triangular crystalline geometry. The quantity $h_0^2/a_S^2$ parameterizes the strain intensity. Importantly, $\theta$ generates different substrate/graphene height profiles, all with the same triangular crystalline structure [see Figs.~\ref{fig:Fig1}(a)--(c)]. Its role is to identify the key geometrical factors affecting band structure and LDOS features. 

Figure~\ref{fig:Fig1}(a) and (d) present representative profiles of $h$ and $\bl{B_p}$ in the $K$ valley resulting from Eq.~(\ref{eq:height}) for $\theta\in[0^\circ,45^\circ]$. To gain an intuitive understanding of the upcoming results, we define a `superlattice unit cell' containing two `superlattice sublattice sites', represented by the two dashed triangles, whose interiors depend on $\theta$. For $\theta = 0^{\circ}$, the profiles contain two equivalent `superlattice sublattices' with identical height distributions in $h$ and opposite signs in $\bl{B_{p}}$. The evolution of these profiles with $\theta$ reveals asymmetries between maximal and minimal field intensities, $\Delta B=|B_{\text{max}}|-|B_{\text{min}}|$, with a  $60^{\circ}$ periodicity, as shown in Fig.~\ref{fig:Fig1}(e) (results for $\theta\in[30^\circ,60^\circ]$ show an inverted triangle profile).  $\Delta B$ is maximal for $\theta=15^{\circ}$ and $45^{\circ}$, and the positive and negative maximal field intensities define two different textures with triangular and kagome-like periodicities [Fig.~\ref{fig:Fig1}(d)]. 
Notably, $\Delta B \neq 0$ signals the breaking of $C_{2z}$ symmetry due to substrate-induced corrugations. The breaking of this symmetry is also manifested in the deformations' spatial profiles: the distribution of $|h|$ in the two triangles in Fig.~\ref{fig:Fig1}(a) are distinct except for $\theta=0^{\circ}$ and $30^{\circ}$.
As will be shown, $C_{2z}$ symmetry breaking is important for gapping out the secondary Dirac points in the mini-Brillouin zone (mBZ), thus giving rise to isolated quasi-flat bands.
Figure~\ref{fig:Fig1}(d) also shows typical equipotential patterns represented by solid curves with that corresponding to $\bl{B_{p}} = 0$ in white. The zero-equipotential lines evolve continuously, from a connected network crisscrossing the whole sample for $\theta = 0^{\circ}$ to a set of disconnected closed curves surrounding regions with asymmetric maximal and minimal  pseudo-field magnitudes. Connected and disconnected regions of $\bl{B_{p}}$ are also observed for other superlattice geometries but with different spatial distributions.

%%%%%%%%%%%%%%%%%%%%%%%%%%%%%%%%%%%%%%%%%%%%%%%%%%%

\begin{figure}[t]
	\center{\includegraphics[width=4.5in]{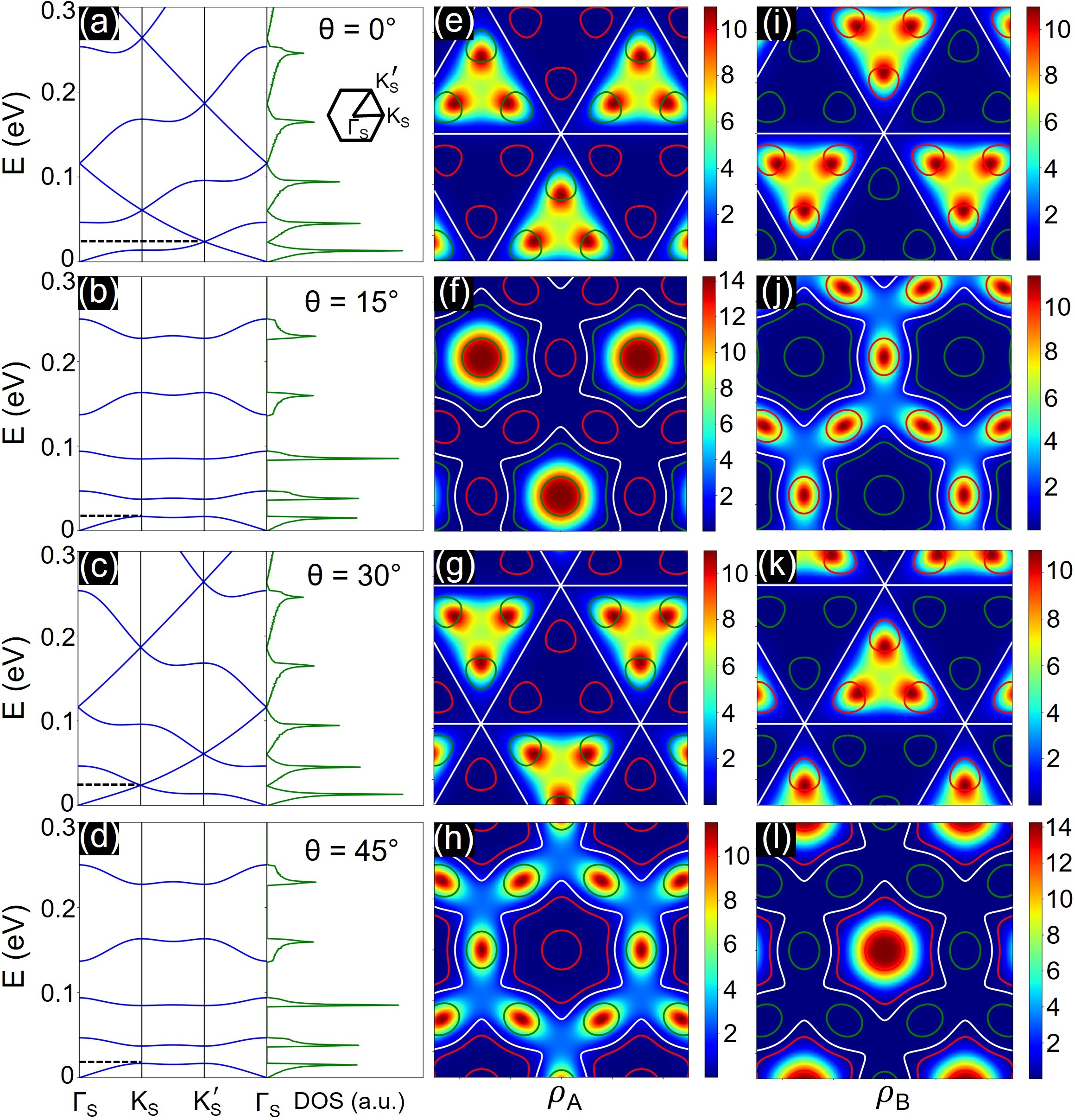}}
	\caption{(a)--(d): Energy bands and total DOS for selected values of $\theta$. LDOS evaluated at energies shown by the dashed lines for sublattice $A$ in panels (e)--(h) and sublattice $B$ in panels (i)--(l). Solid lines indicate pseudo-magnetic field equipotentials. Panels with $C_{2z}$ symmetric profiles: (e) \& (i) and (g) \& (k): $\bl{B_{p}} = - 300$ T (green), $300$ T (red). Panels with $C_{2z}$ broken profiles: (f) \& (j): $\bl{B_{p}} = - 200$ T (green), $350$ T (red), (h) \& (l):  $\bl{B_{p}} = - 350$ T (green), $200$ T (red). White curves denote $\bl{B_{p}} = 0$ in all panels. $h_0 = 0.6$~nm, $a_S = 14$~nm.}
	\label{fig:Fig2}
\end{figure}

\begin{figure}[H]
	\includegraphics[width=5in]{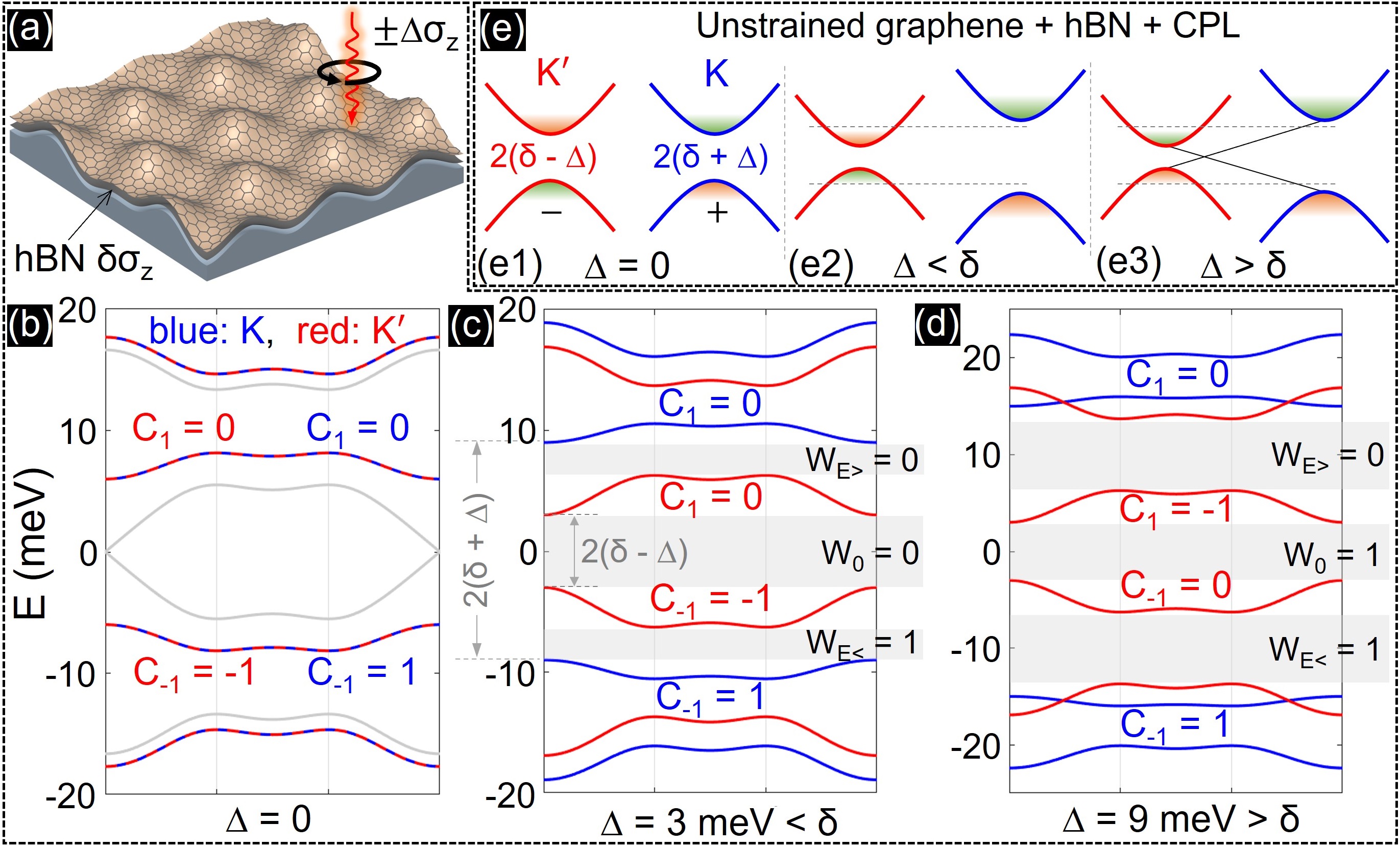}
	\caption{(a) Schematics of graphene on a substrate with an hBN spacer (black surface) and CPL illumination. hBN and CPL introduce staggered potentials $\delta\sigma_{z}$ and $\pm\Delta\sigma_{z}$ respectively. (b--d) Energy bands of the $K$ (blue) and $K'$ (red) valley with hBN ($\delta\equiv6$ meV) and CPL of different intensities: $\Delta=0$ (b), $\Delta=3$ meV (c), $\Delta=9$ meV (d). The k-space high symmetry path in the mBZ is the same as that in Fig.~\ref{fig:Fig2}. Chern numbers of the first conduction and valence bands in each valley are explicitly shown, the other bands shown have $C=0$. Grey curves in (b) are energy bands when $\delta=\Delta=0$. Grey regions in (c) and (d) highlight some gaps and the corresponding winding numbers. (e) Schematic low-energy bands of unstrained graphene under the same conditions as (b)--(d). A gap of size $2(\delta\pm\Delta)$ is opened in the two valleys. The orange/green shadows denote positive/negative Berry curvature hot spots, which are strongly localized near the band edges when the gap is small. Each Berry curvature hot spot can contribute a Hall conductance $\mathcal{O}(\pm\frac{e^2}{2h})$ per spin. The horizontal grey dashed lines in (e2) and (e3) represent Fermi levels within the energy window of $(-\delta-\Delta,\,\delta+\Delta)$, which serve as counterparts of the Fermi levels in the upper and lower grey shaded gaps of (c) and (d). When the Fermi level is tuned from $E=0$ to either dashed line, the Hall conductance is expected to change by the amount $\mathcal{O}(\frac{e^2}{2h})$ per spin, which is different from the situations in (c) and (d). The black solid lines in (e3) represent chiral edge states that appear in a ribbon. Here $h_0=0.8$ nm, $a_S=20$ nm, $\theta=15^\circ$.}
	\label{fig:Fig3}
\end{figure}

Next we look at the electronic properties, with a focus on the $K$ valley (see SI for results in the $K'$ valley).
The $\theta$-dependent deformation profiles give rise to distinct patterns in band structures, total DOS, and LDOS at graphene sublattice sites $A$ and $B$, as shown in Fig.~\ref{fig:Fig2}. Only the conduction bands are shown as $\bl{A_p}$ preserves chiral symmetry, which yields mirror symmetric bands and protects graphene's Dirac point at $\Gamma_S$ [e.g., grey curves in Fig.~\ref{fig:Fig3}(b)]. A comparison between LDOS patterns [Figs.~\ref{fig:Fig2}(e)--(l)] and the $\bl{B_{p}}$ profiles [Fig.~\ref{fig:Fig1}(d)] reveals that LDOS of sublattice $A$ ($B$) localizes at regions of negative (positive) $\bl{B_{p}}$ values. Intuitively, these two distributions can be visualized as the alignment between the sublattice pseudo-spin up (down) and the local negative (positive) field distribution to lower the energy, i.e., akin to a pseudo-Zeeman effect~\cite{Manes2013,Georgi2017,Snyman2009}, a feature independent of the periodic structure considered.

As $\theta$ changes, one can distinguish two regimes depending on $\bl{B_{p}}$ profiles with or without $C_{2z}$ symmetry. Values of $\theta$ that preserve the symmetry produce gapless spectra, as shown in Figs.~\ref{fig:Fig2}(a) and (c), with secondary Dirac points at $K_S$, $K'_S$, and $\Gamma_S$ at $E\ne0$ connecting all the bands.
The gapless spectra are accompanied by sublattice LDOS distributions with same magnitudes as presented in panels (e) \& (i) and (g) \& (k). In contrast, values of $\theta$ that break the $C_{2z}$ symmetry render gapped spectra and distinct sublattice LDOS distributions, as shown in the second and fourth rows of Fig.~\ref{fig:Fig2}. 
These bands and the associated DOS peaks, unlike Landau levels, show varying inter-level separations without evident scaling behavior with $B_p$’s magnitude.
The gap opening at the secondary Dirac points occurs for all values of $\theta \neq 0^{\circ}$ or $30^{\circ}$, and can be associated with the macroscopically inequivalent `superlattice sublattice' occupations, akin to the effect of alignment of graphene on hBN that breaks graphene's sublattice symmetry. 
Figure~\ref{fig:Fig2}(f) \& (j) and (h) \& (l) show examples for $\theta = 15^{\circ}$ and $45^{\circ}$ with maximal gap openings accompanied by maximal sublattice LDOS contrast [see Fig.~\ref{fig:Fig1}(e)].  Variations in other parameters, i.e., height $h_0$, superlattice period $a_S$, or lattice geometry, do not modify the existence of these two regimes (but the band widths and gaps sizes might vary among different superlattice geometries with similar geometric parameters, see SI), thus confirming that $C_{2z}$ symmetry breaking, or equivalently the sublattice asymmetry, is at the origin of gap openings~\cite{Phong2021a}. Therefore, patterned substrates with pronounced $C_{2z}$ breaking profiles [e.g., Figs.~\ref{fig:Fig1}(b) and (c)] are most desirable for achieving isolated quasi-flat bands. 

Experimentally, properties of the band structure (e.g., band width, gap size) and the LDOS textures can be probed with scanning tunneling spectroscopy~\cite{Mao2020}. The effects of the substrates in such measurements are expected to be either negligible or easily isolated from the signatures of graphene, provided the two materials do not hybridize. One common effect of the substrate on graphene is doping.~\cite{Mao2020} Such effect can be easily calibrated by performing similar measurements in graphene on a flat substrate made of the same material. These control experiments not only can be employed to identify the new features introduced by the superlattice patterning, but also help to tell whether the hybridization between graphene and the substrate is negligible.

%%%%%%%%%%%%%%%%%%%%%%%%%%%%%%%%%%%%%%%%%%%%%%%%%%

The presence of pseudo-magnetic fields implies interesting topological properties in periodically strained graphene. The first conduction and valence bands, which are degenerate at the Dirac point, have a total Chern number $C_{1}+C_{-1}=1 ~ (-1)$ for $\theta=15^\circ$ ($45^\circ$) in the $K$ valley [e.g., Fig.~\ref{fig:Fig3}(b)]. The Chern numbers flip sign in the $K'$ valley due to TRS.
The other energy bands shown throughout this work have $C = 0$. These Chern numbers stay invariant for deformations defined by $0^\circ<\theta<30^\circ$ and $30^\circ<\theta<60^\circ$ as the band gaps do not close within these ranges [e.g., Fig.~\ref{fig:Fig1}(e)]. 
Importantly, we find that the two sublattices of graphene, whose LDOS are spatially separated with distinct profiles (e.g., Fig.~\ref{fig:Fig2}), have different topological properties: the sublattice with extended LDOS [e.g., Fig.~\ref{fig:Fig2}(j) or (h)] contributes to the nontrivial topology; while the sublattice with localized LDOS [e.g., Fig.~\ref{fig:Fig2}(f) or (l)] renders $C=0$. Similar features are also present in other superlattice geometries (see SI). These results suggest that patterned substrates producing continuous strain distributions with fully connected $\bl{B_p}$ profiles (thus extended LDOS features) are crucial for achieving topological flat bands.

Experimental signatures of the band topology in periodically strained graphene are hard to obtain. A recent theoretical study found counterflowing in-gap boundary modes from the two valleys that reside on the same edge in a zigzag ribbon with periodic strain~\cite{Phong2021a}. The presence of such boundary modes requires high-quality samples with zigzag edges to suppress intervalley scattering. Moreover, as no charge Hall response exists due to TRS, the experimental probe of these boundary modes is nontrivial. 
Previous works have suggested that the valley Hall phase can contribute to nonlocal transport signals~\cite{VHENatPhys2015a,VHENatPhys2015b}, which could be used to reveal the underlying band topology, but such measurements in strained graphene are lacking.
In the following, we propose an optical scheme to probe these unique topological properties. We show that chiral edge states that contribute to charge Hall responses can emerge in the superlattice energy gaps, thus providing unique experimental fingerprints of strain-induced band topology.

Our scheme is composed of two ingredients [Fig.~\ref{fig:Fig3}(a)]: (i) A thin hBN spacer between graphene and the substrate~\cite{hBNSpacerImamoglu2020,hBNSpacerNatPhotonics2022}, which preserves graphene's deformation profile but breaks its Dirac point degeneracy by a valley-symmetric staggered potential $H_{\text{hBN}}=\delta\sigma_{z}$; (ii) A circularly polarized light (CPL), whose role is to break TRS. Under these conditions, robust chiral edge states that contribute to charge Hall responses are allowed to emerge.
We consider a left CPL, whose vector potential reads $\bl{A}_{\text{CPL}}=A_0(\cos\omega t,\,-\sin\omega t)$. In the high-frequency limit $\hbar\omega\gg evA_0$, and at low-energies $E\ll\hbar\omega$ of experimental interest, electrons interact with the light mainly via off-resonant processes~\cite{FloquetReviewRudnerNatRevPhys2020,FloquetReviewOkaAnnuRev2019,FloquetReviewAdvPhys2015}. The effects of the CPL on the electronic properties of graphene then can be captured by a valley-antisymmetric staggered potential $H_{\text{CPL}}=\tau \Delta \sigma_{z}$, where $\Delta=(evA_0)^2/(\hbar\omega)$. This approach is well established in pristine monolayers, and recently has been shown to be valid also in moir\'e superlattices~\cite{FloquetTBGRubio2019,FloquetTBGGil2020,FloquetTBGBabak2020,FloquetTDBGMartin2020}. For a CPL with $\hbar\omega=3$ eV and electric field intensity $\omega A_0=6$ MV/cm, one finds $\Delta\approx6$ meV.
With both ingredients, the low-energy Hamiltonian of the system [Fig.~\ref{fig:Fig3}(a)] becomes 
\begin{equation}
H_{\tau}(\text{hBN},\text{CPL}) = v\bl{\sigma}_{\tau}\cdot(\bl{p}+\tau e\bl{A_p})+(\delta+\tau\Delta)\sigma_{z}.
\end{equation}

We use the substrate defined by $\theta=15^\circ$ to illustrate the control processes in the following [Figs.~\ref{fig:Fig3}(b)--(d)]. Results for $\theta=45^\circ$ and right CPL can be found in SI. We consider superlattices that have larger periods, e.g., $\gtrsim20$ nm, which are consistent with current substrate engineering capabilities, and produce narrower bands with smaller strain [cf. Fig.~\ref{fig:Fig2}(b)].
The value of $\delta$ depends on the alignment between graphene and hBN, which can reach $\mathcal{O}(10)$~meV~\cite{GrapheneGaphBNScience2013,hBNGapNanoLett2018}. Here we use $\delta=6$ meV, a small value that assumes certain misalignment between graphene and hBN on the patterned substrate, and tune the value of $\Delta$, e.g., by changing the intensity of the CPL.
As the net staggered potential in the two valleys are $(\delta+\Delta)\sigma_{z}$ vs $(\delta-\Delta)\sigma_{z}$, the Dirac point degeneracy of graphene is broken by a gap of size $2(\delta+\Delta)$ and $2(\delta-\Delta)$ at the $K$ and $K'$ valleys, respectively. Since $\delta$ and $\Delta$ compete in the $K'$ valley, there exist different regimes depending on their relative size.

Figure~\ref{fig:Fig3}(b) shows the energy bands of the superlattice with their Chern numbers in the absence of CPL ($\Delta=0$). By comparing the red/blue vs grey curves, one identifies that the major effects of a staggered potential (here $\delta\sigma_{z}$ from hBN) are to break the Dirac point degeneracy and reduce the bandwidths of the two central bands. The other bands are minimally affected, exhibiting slight energy shifts only.
Regarding the band topology, a staggered potential renders the two isolated conduction and valence bands with trivial and nontrivial topology, the details of which depend on the sign of the potential. For example, in Fig.~\ref{fig:Fig3}(b), the first valence (conduction) band is topologically nontrivial (trivial). This can be intuitively understood by noticing that the positive staggered potential polarizes the states in the valence (conduction) band towards the B (A) sublattice with a delocalized (localized) LDOS [e.g., Figs.~\ref{fig:Fig2}(j) and (f)]. However, because the total Chern number of the two valleys vanishes due to TRS, there are no topologically protected chiral edge states that could contribute to charge Hall responses. The situation becomes drastically different when the CPL is introduced.

Figure~\ref{fig:Fig3}(c) shows the results in the regime of $0<\Delta<\delta$. Compared to the previous case, the Chern number of each band remains unchanged, but TRS is broken, which is manifested by the different structures of the bands in the two valleys. These features can be easily understood as the staggered potentials in the two valleys are both positive but have different magnitudes. As TRS is broken, chiral edge states --manifestations of the band topology-- might emerge in the energy gaps, depending on the value of the corresponding topological invariant.~\cite{FloquetReviewRudnerNatRevPhys2020,FloquetWindingNumberPRX2013,TopologicalInsulatorRMP2010,TopologicalInsulatorRMP2011}
In a driven system, this topological invariant is the winding number $W_E$ inside a gap centered at energy $E$, which characterizes the number of chiral edge states inside this gap~\cite{FloquetWindingNumberPRX2013}. The winding number in the zero-energy gap in Fig.~\ref{fig:Fig3}(c), $W_0$, can be obtained by first considering the case without strain~\cite{FloquetTBGGil2020}. In the unstrained case, and the high-frequency limit, graphene with gaps of $2(\delta\pm\Delta)>0$ at the two Dirac points [e.g., Fig.~\ref{fig:Fig3}(e2)] yields $W_{0}=\frac{1}{2}\,\text{sgn}(\delta+\Delta)-\frac{1}{2}\,\text{sgn}(\delta-\Delta)=0$~\cite{Haldane1988,FloquetGraphenehBNPRB2014}. As these two gaps do not close after strain is added, one finds $W_0=0$ in Fig.~\ref{fig:Fig3}(c) for the strained case. 
Winding numbers in the other gaps can be obtained from the Chern numbers of the bands by employing the relation $W_{\mathcal{E}}=W_{\varepsilon}+\sum C_n$, where $\mathcal{E}>\varepsilon$, and the sum is over all the bands between the gaps centered at $\varepsilon$ and $\mathcal{E}$~\cite{FloquetWindingNumberPRX2013}.
In Fig.~\ref{fig:Fig3}(c), as the two red bands in the middle show $C_1=0$ and $C_{-1}=-1$, one finds $W_{E_>}=0$ and $W_{E_<}=1$ (per spin) in the upper and lower gaps, respectively. Consequently, in the regime of $0<\Delta<\delta$, the presence of chiral edge states in one of the superlattice gaps, and the anomalous Hall effect with nearly quantized Hall conductance (see Refs.~\citenum{FloquetHallConductanceFoaPRL2014,FloquetHallConductancePRB2015,FloquetAHEexptNatPhys2020,FloquetGrapheneRubioPRB2019} for more discussions on values of conductance in driven systems), serve as evidence of the strain-induced nontrivial band topology.

The uniqueness of strain is also clear by comparing to the unstrained case. Figure~\ref{fig:Fig3}(e2) shows schematic low-energy bands and Berry curvature hot spots (orange/green shadows) in an unstrained graphene under the same conditions. The strain-induced superlattice gaps, thus the in-gap chiral edge states, are obviously absent. Additionally, the Hall conductance significantly deviates from $\frac{e^2}{h}$ (per spin) in the absence of strain when the Fermi level is in the window of $(-\delta-\Delta,\,\delta+\Delta)$ [e.g., grey dashed lines in Fig.~\ref{fig:Fig3}(e2)] as the Berry curvature hot spots can only contribute  $\mathcal{O}(\pm\frac{e^2}{2h})$.

Figure~\ref{fig:Fig3}(d) shows the results in the regime of $\Delta>\delta$. As $\Delta$ is tuned from below $\delta$ to above, the conduction and valence bands in the $K$ valley become more separated and their Chern numbers are unaffected [blue curves, Figs.~\ref{fig:Fig3}(c) vs (d)]. In contrast, a topological phase transition occurs in the $K'$ valley due to gap closing at the Dirac point (at $\Delta=\delta$, the grey curves in Fig.~\ref{fig:Fig3}(b) are reproduced) and reopening (when $\Delta>\delta$), during which the first conduction and valence bands exchange Chern numbers [red curves, Figs.~\ref{fig:Fig3}(c) vs (d)]. Additionally, the winding number in the gap centered at $E=0$ becomes $W_0=1$ as the staggered potential in the $K'$ valley turns negative when $\Delta>\delta$~\cite{Haldane1988}. Note however that chiral edge states also exist in the gap at $E=0$ in unstrained graphene in this regime [Fig.~\ref{fig:Fig3}(e3)]~\cite{FloquetGraphenehBNPRB2014,FloquetHallConductanceFoaPRL2014,FloquetHallConductancePRB2015}. The strained and unstrained cases are distinguishable by the evolution of the chiral edge states and Hall conductance at different energies within $(-\delta-\Delta,\,\delta+\Delta)$.
Specifically, in Fig.~\ref{fig:Fig3}(d), the chiral edge states disappear in the upper gap as $W_{E_>}=0$ (due to $C_1=-1$ in red), while they are preserved in the lower gap as $W_{E_<}=1$ (due to $C_{-1}=0$ in red). 
The complete suppression of Hall conductance in the upper gap in Fig.~\ref{fig:Fig3}(d) is a consequence of strain-induced nontrivial band topology in the regime of $\Delta>\delta$, and it cannot be observed in unstrained graphene within a similar energy window [see Fig.~\ref{fig:Fig3}(e3)].
In fact, the chiral edge states with their associated Hall conductance in the lower gap of Fig.~\ref{fig:Fig3}(d) cannot be observed either in unstrained graphene.

In summary, we have shown that patterned substrates with pronounced $C_{2z}$ symmetry breaking profiles are necessary to obtain isolated quasi-flat bands. Connected landscapes in the pseudo-magnetic field are important to imprint nontrivial topology to the bands. Band flatness, topology, and presence/absence of chiral edge states can be controlled with an hBN spacer and CPL. Our results not only offer important guidance on designing substrates for tailored electronic and topological properties, but also provide robust experimental signatures for identifying the strain-induced band topology by using CPL. 
The chiral edge states and results of Hall conductance for Fermi levels inside the bulk gaps are expected to be robust against strain fluctuations and other sources of random disorder, provided the energy gaps of interest are not closed. Notice that, in contrast, the boundary modes in the absence of CPL~\cite{Phong2021a} are fragile against such perturbations.
The obtained topological quasi-flat bands might be promising platforms for exploring correlation-driven phenomena~\cite{Gao2022}. Effects of an external magnetic field~\cite{LinHeStrainAndRealMagneticFieldPRL2020} on the strained superlattice is also interesting to explore in the future.

\textbf{Supporting Information}. Details of strain pseudo-vector potential and scalar field, results of $K'$ valley, effects of the scalar potential, impacts of in-plane displacements, results for other geometric parameters, extra figures for optical control, and results for other superlattice geometries.

\begin{acknowledgement}
We acknowledge discussions with O. Avalos-Ovando, D. Faria, S. E. Ulloa, K. I. Ingersent, M. M. Asmar, and B. Fu. Md.\,T.\,M.  was partially supported by NQPI, Ohio University. Portions of this work were completed at NBI (KU) and the Physics Department at DTU (Denmark) under support from Otto M{\o}nsteds and NORDEA foundations (NS).
\end{acknowledgement}

\bibliography{SubstrateRef}

\end{document}